\documentclass[%
 preprint,
superscriptaddress,
 amsmath,amssymb,
 aps,
prl,
]{revtex4-1}

\usepackage{graphicx}
\usepackage{dcolumn}
\usepackage{bm}
\usepackage{gensymb}
\usepackage{upgreek}

\usepackage{amsfonts}
\usepackage{booktabs}

\begin{document}

\def\SiO2{$\mathrm{SiO_2}$}
\def\bq{``}
\def\NbO2{$\mathrm{NbO_2}$}
\def\Nb2O5{$\mathrm{Nb_2O_5}$}
\def\Rs{$\mathrm{R_S}$}
\title{Growth and Characterization of Polycrystalline \textbf{NbO}$_2$ Thin Films on Crystalline and Amorphous Substrates}

\author{Ali Fakih}
\affiliation{%
Institut de Min\'{e}ralogie, de Physique des Mat\'{e}riaux et de Cosmochimie,  Sorbonne Universit\'{e}, UMR CNRS 7590, MNHN, 4 Place Jussieu, F-75005 Paris, France
}%

\author{Johan Biscaras}
\email{johan.biscaras@sorbonne-universite.fr}
\affiliation{%
Institut de Min\'{e}ralogie, de Physique des Mat\'{e}riaux et de Cosmochimie,  Sorbonne Universit\'{e}, UMR CNRS 7590, MNHN, 4 Place Jussieu, F-75005 Paris, France
}%

\author{Abhay Shukla}
\email{abhay.shukla@sorbonne-universite.fr}
\affiliation{%
Institut de Min\'{e}ralogie, de Physique des Mat\'{e}riaux et de Cosmochimie,  Sorbonne Universit\'{e}, UMR CNRS 7590, MNHN, 4 Place Jussieu, F-75005 Paris, France
}%

\begin{abstract}
\NbO2 is a potential material for nanometric memristor devices, both in the amorphous and the crystalline form. We fabricated \NbO2 thin films using  RF-magnetron sputtering  from a stoichiometric target. The as-deposited films were amorphous regardless of the sputtering parameters. Post deposition vacuum annealing of the films was necessary to achieve crystallinity. A high degree of crystallinity was obtained by optimizing annealing duration and temperature. The resistivity of the material increases as it undergoes a structural transition from amorphous to crystalline with the crystalline films being one order of magnitude more resistive.\\ 

\begin{description}
\item[DOI] 
\hspace{8cm} 
\textbf{PACS numbers:}
\end{description}
\end{abstract}

\maketitle


\section{Introduction} 
Niobium dioxide \NbO2 exhibits an insulator-metal transition at about 800 $\degree$C which transforms the low temperature insulating phase with a room temperature resistivity $\rho$ of $2\times 10^2\, \Omega$.m \cite{nakao} into a metal with a resistivity of $1.4\times 10^{-3}\, \Omega$.cm \cite{resistivity-sakata}. This phase transition is accompanied by a structural transition from a distorted rutile phase space group (I4$_1$/a) to a rutile phase space group (P4$_2$/mnm) \cite{structure1,structure2,structure3,structure4}. Whether this phase transition is of the kind found in its parent metal oxide VO$_2$ at a much lower temperature is an open question, but the high temperature associated with it, and thus the stability of the low temperature phase has motivated several groups to study \NbO2 based devices \cite{device1,device2,device3,device4,device5,device6}. 
These devices also show a sizeable change in resistance on the application of current, or more accurately, a current controlled negative differential resistance regime \cite{device6}. While devices often use variable stoichiometry NbO$_x$ (with x varying from 2 to 2.5) and amorphous as-deposited phases which eventually crystallize with temperature or voltage \cite {device4}, it is desirable to directly grow the \NbO2 phase, both amorphous and crystalline, on a variety of substrates, using a single, reproducible method. Numerous works have reported the growth of \NbO2 films using evaporation \cite{zhao,mbe2,mbe3,mbe4}, deposition \cite{nakao,wang,dep} or sputtering techniques \cite{music,sputtering2,sputtering3,sputtering4,sputtering5,sputtering6,sputtering7,sputtering8}    with either in-situ or post-deposition oxidation and crystallization on various substrates ranging from glass to single crystal silicon.\\

The targets used have been variously Nb or stoichiometric \NbO2 with oxygen or argon partial pressure during deposition. Post deposition annealing, eventually in an oxidizing atmosphere has also been used. For example, \NbO2 thin films with nanoslice structures were obtained on silicon substrate by using  reactive magnetron sputtering on a niobium target in Ar/O$_2$ atmosphere \cite{music}. Epitaxial \NbO2 thin films were fabricated on Al$_2$O$_3$ substrates using reactive bias target ion beam deposition \cite{wang}. Recently Nakao et al.\cite{nakao} demonstrated that \NbO2 films can be grown with pulsed laser deposition in a partial pressure of oxygen using a stoichiometric target. To  attain crystallinity their samples were  annealed after deposition in vacuum  at a temperature of 600$\degree$C. However, it was also  noted that the resistivity obtained for their samples remained significantly below the  resistivity of bulk \NbO2 . Amorphous \NbO2 has a lower resistivity than crystalline \NbO2, and thus low resistivity in thin films samples implies imperfect crystallization or eventually a deficient oxygen stoichiometry. For reference, bulk resistivity of \NbO2 is  $1\times 10^4\,\Omega$.cm  \cite{nakao} and that of  \Nb2O5 is $2 \times \, 10^7\, \Omega$.cm \cite{resistivity-nb2o5}. In between these two phases \NbO2 (Nb charge state $4+$) and \Nb2O5 (Nb charge $5+$) several non-stoichiometric, polymorphic or meta-stable phases exist which complicates fabrication of pure phases, especially that of \NbO2 \cite{zhao,mbe2,mbe3,mbe4,nakao,wang,dep,music,sputtering2,sputtering3,sputtering4,sputtering5,sputtering6,sputtering7,sputtering8,difficulty}.

\section{Materials and Methods} 
To directly deposit stoichiometric \NbO2 on the substrate, we chose RF magnetron sputtering avoiding partial oxygen pressure during deposition. We decided to concentrate on adequate conditions for deposition  such as sputtering rate, substrate temperature and Ar partial pressure to obtain thin films with the good stoichiometry,  followed by  vacuum annealing for crystallization.\\

\begin{table}[h]
\centering
\begin{tabular}{cccc} \toprule

\hline

    {Parameter} & {Target Etching } & {Substrate Etching} & {Deposition}  \\ \hline
  Ar pressure (mbar)  & 0.5 & 0.5 & 0.5  \\
   power (Watts)  & 100  & 225 & 225  \\
   voltage (V)  & 465  & 480 & 500  \\
  target-substrate distance (cm)  & 6.5  & 6.5 & 6.5   \\
  substrate temperature ($\mathrm{\degree C}$) & RT &  RT & 550    \\
    time (min)  & 2  & 3 & 4    \\ \hline
\end{tabular}
\caption{Optimized sputtering parameters for stoichiometric \NbO2 thin films.}
\label{parameters1}
\end{table}

The stoichiometric target  was obtained from Edgetech  Industries LLC. It is important to verify that the target presents a uniform dark blue-grey colour characteristic of the \NbO2 phase. A light coloured target is a sign of oxidation during the sintering phase of the target even if the powder \,  precursor \,  is \, stoichiometric \NbO2. The \NbO2 phase of the chosen target was verified by Raman spectroscopy and X-Ray diffraction (XRD) before mounting in the sputtering chamber. Nominal stoichiometry for the target does not imply phase purity and stoichiometry of the deposited film and may be the reason why oxygen or argon, and in some cases hydrogen \cite{annealing1982} partial pressure has been necessary in earlier experiments. However oxidation or reduction during film deposition is a delicate adjustment which can also influence deposition dynamics.\\

An important aspect of our methodology was to ensure a similar fabrication process for films on three different substrates, soda-lime glass, single crystal silicon   and single crystal silicon topped by a 285 nm layer of poly-crystalline oxide. Of these, the glass substrate constrains annealing to temperatures below 600$\degree$ C, while the other substrates may be annealed at higher temperatures. The parameters varied to optimize the quality of the deposited films were total magnetron power, the process gas (Ar) pressure, sample target distance which control the sputtering rate and finally substrate temperature which can alter deposition dynamics and notably help crystallization.  Several depositions were made with the three kinds of substrates by varying these parameters. We observed that the dominant parameter controlling the film quality was the Ar pressure. The final optimal  parameters for each of the substrates  are collected in table \ref{parameters1}.\\

The Raman Spectroscopy measurements were performed using an Xplora Raman spectrometer (HORIBA Jobin-Yvon) with a monochromatic laser of  532 nm wave length, and  the XRD instrument used  was  Panalytical X’pert Pro MPD diffractometer with a  cobalt cathode ray tube as an x-ray source having $\lambda_{\mathrm{Co}}= 1.790$ \AA.   XRD experiments were measured in two different configurations: (1) Bragg Brentano (BB) and (2) grazing incidence (GI) with an angle fixed at $1\degree$. In thin films (some samples were only 30 nm thick) the grazing incidence mode helps to maximize the signal from the sample and minimize the signal from the substrate. Both Raman scattering and XRD were used to check the phase of the poly-crystalline films. Finally, we  optimized post-annealing temperature for each substrate and characterized sample quality using optical micrographs, XRD, Raman scattering and transport measurements.

\section{Results and Discussion}
Using the parameters presented in table \ref{parameters1}, amorphous films were obtained on all three substrates regardless of the substrate temperature as verified by both Raman spectroscopy and XRD. Trace crystalline structure was obtained on the Si and SiO$_2$/Si. The deposition rate was about 33 nm/min as measured by AFM. After the deposition, an annealing procedure was then implemented to induce crystallization \, in \,  the \,  samples. Keeping in mind possible oxidation to the unwanted \Nb2O5 phase in an oxidizing environment, we used  vacuum annealing of the as-deposited films  which  yielded large area poly-crystalline films.\\

\begin{figure}[t]
\hspace*{-0.2cm}
\centering
\includegraphics[scale=0.45]{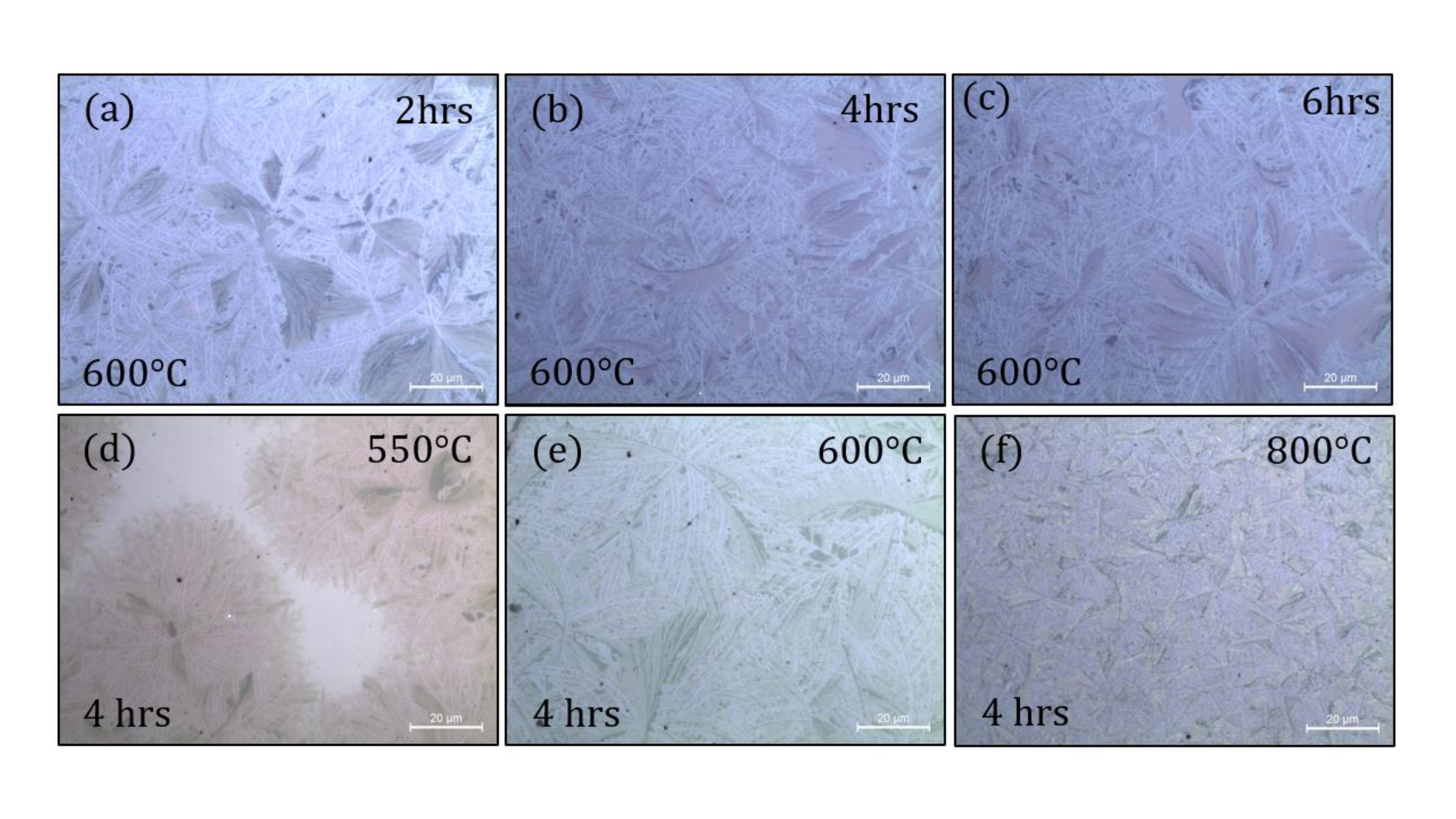}
\caption{Optical images of 330 nm \NbO2 films. Films on glass substrates annealed at different times (a) 2hrs (b) 4hrs (c) 6hrs.  Films on Si substrates annealed at different temperatures (d) 550 $\degree$, non crystalline regions remain in the film (e) 600$\degree$C (f) 800$\degree$C.}
\label{parameters2}
\end{figure}

A typical branching microstructure (see Figure \ref{parameters2}) appears on crystallization and is identified as \NbO2 with micro Raman spectroscopy. Optical inspection shows that microstructure size and density changes  with annealing time (Figure \ref{parameters2} (a), (b) and (c), for \NbO2 films on glass). The annealing temperature plays a crucial role in crystallization since below about 525$\degree$C crystallization does not occur. On increasing the annealing temperature (Figure \ref{parameters2} (d), (e) and (f), \NbO2 film on Si) we notice that crystallization is not complete at 550$\degree$C (Figures \ref{parameters2} (d)), while above 600$\degree$C (Figures \ref{parameters2} (e) and  (f)) microstructure size and density appear homogeneous.\\ 
%
%
%
%
%

\begin{figure*}[!htb]
\hspace*{-1 cm}
\centering
\includegraphics[scale=0.40]{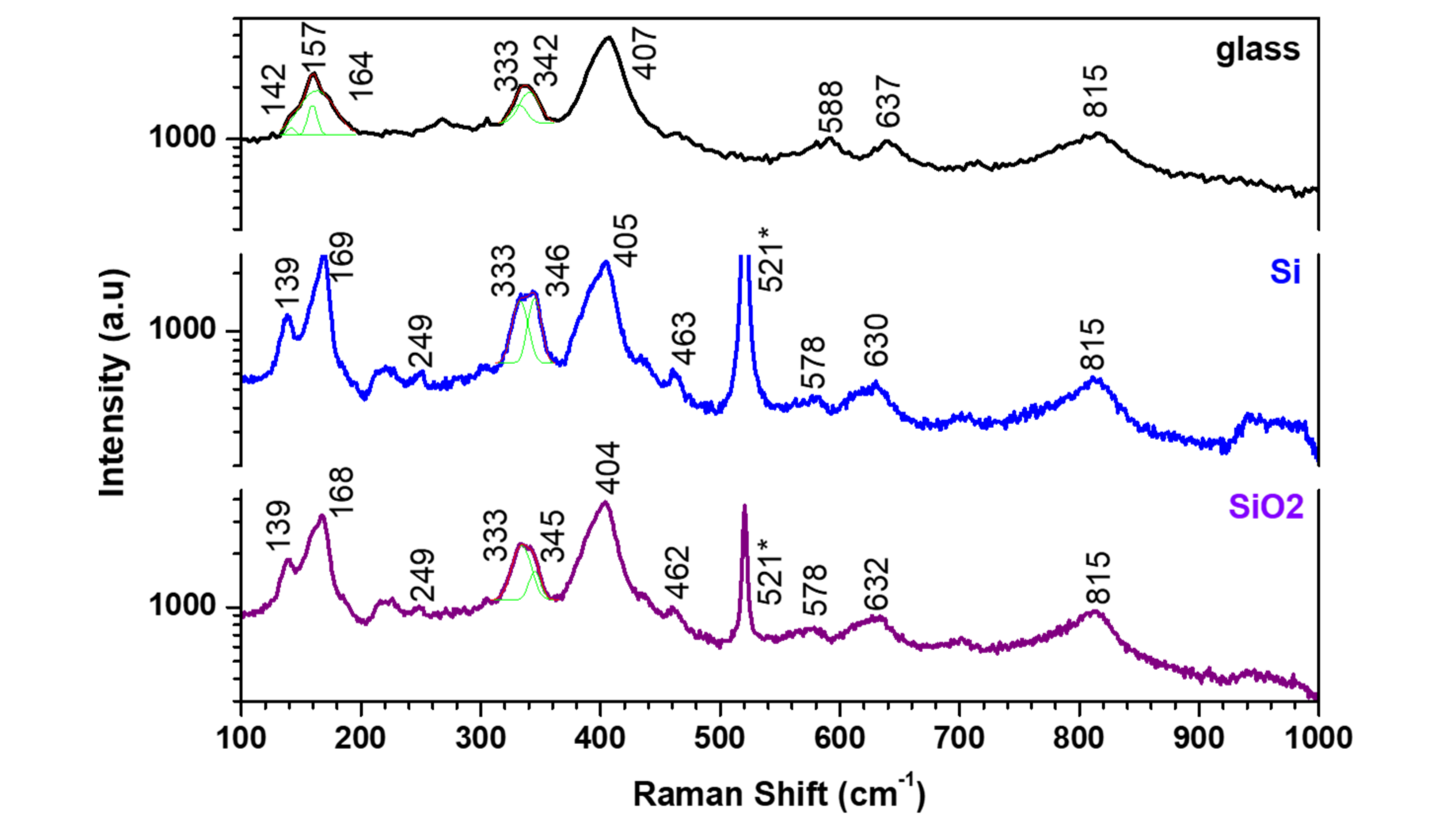}
\caption{The Raman Spectra of \NbO2 films deposited on different substrates: glass, Si $<100>$ , and SiO$_2$. The 521 $cm^{-1}$ peaks come from the silicon substrate.  }
\label{raman-paper}
\end{figure*}

Raman  spectra of the \NbO2 films deposited on different substrates (glass, Si, and SiO$_2$) are shown in Figure \ref{raman-paper}. These spectra match those of  nanostructured \NbO2 thin films prepared by different methods \cite{zhao,wang,wong}. \\

%
%
%
%
%
%
%
%
%
%
%
%
%
\begin{figure*}[!htb]
\hspace*{-1.8cm}
\centering
\includegraphics[scale=0.6]{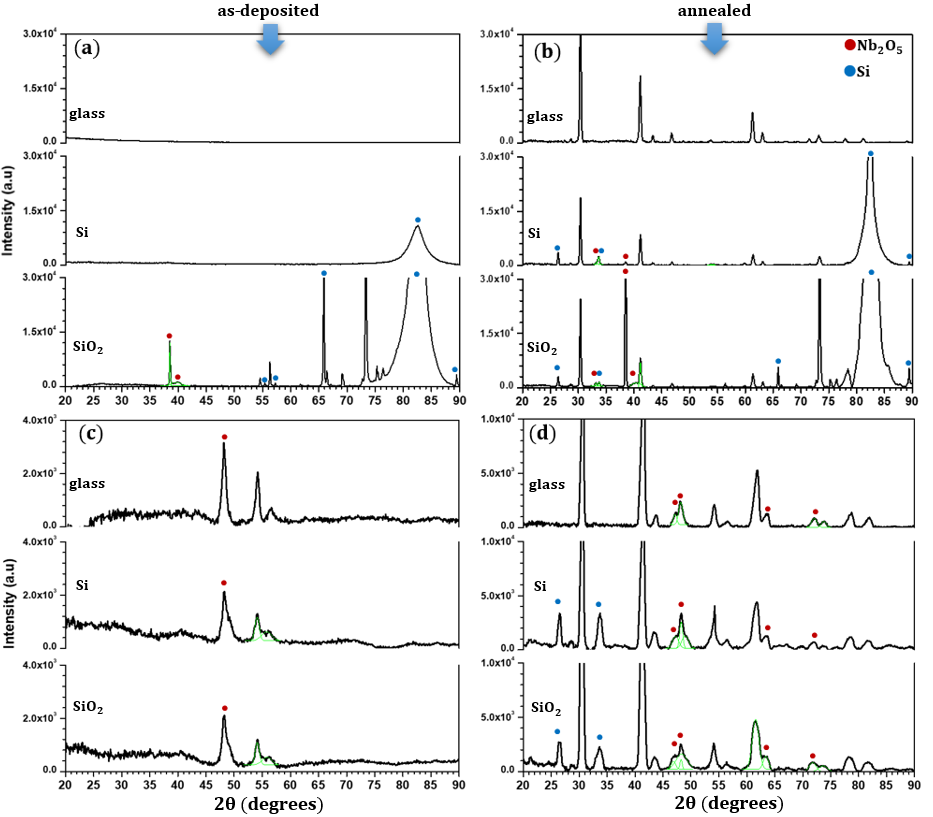}
\caption{X-ray diffractograms of \NbO2 films on different substrates: glass, Si and SiO$_2$ before and after annealing them. Figures (a) and (b) show the results of experiments done in the  Bragg-Brentano configuration, while figures (c) and (d) show the results of experiments done in the grazing incidence geometry. }
\label{xray-paper}
\end{figure*} 
XRD  measurements are shown in the standard Bragg Brentano (BB) geometry (Figures \ref{xray-paper} (a) and (b)) and the  Grazing Incidence (GI) geometry (Figures \ref{xray-paper} (c) and (d)). 
In the as-deposited samples on the three substrates (Figures \ref{xray-paper} (a) and (b)) only the films on the SiO$_2$/Si substrate show signs of crystallized phases of both \NbO2 and \Nb2O5 \cite{nbo24,nbo25,nbo28,nbo21,nb2o52,nb2o53,nb2o55,nb2o56,si1}. Trace \NbO2 or \Nb2O5 peaks on all three substrates are visible in the grazing incidence spectra. After vacuum annealing, all the three samples show the crystallized \NbO2 phase. Trace contributions from crystallized \Nb2O5 are visible in the GI spectra. 
It is well-known \cite{sputtering6} that exposure of \NbO2 thin films to air can lead to a surface \Nb2O5 layer. Nevertheless, the quality of the films can be attested by the dominant \NbO2 peaks.\\

The grain size of the \NbO2 crystallites composing the films was estimated to be between 20 and 40 nm from the peak widths of diffractograms measured in BB geometry using the Scherrer equation:
$$\mathrm{ \tau = \frac{K \,\lambda}{\upbeta \,cos\uptheta}}$$

where $\tau$ is the mean size of the crystallites which may be smaller or equal to the grain size, K is a dimensionless shape factor with a typical value of 0.9, $\lambda$ is the x-ray wave length, $\upbeta$ is the peak broadening at half the maximum intensity (FWHM) and $\uptheta$ is the Bragg angle.\\

\begin{figure}[!htb]
\hspace*{-1.0cm}
\centering
\includegraphics[scale=0.45]{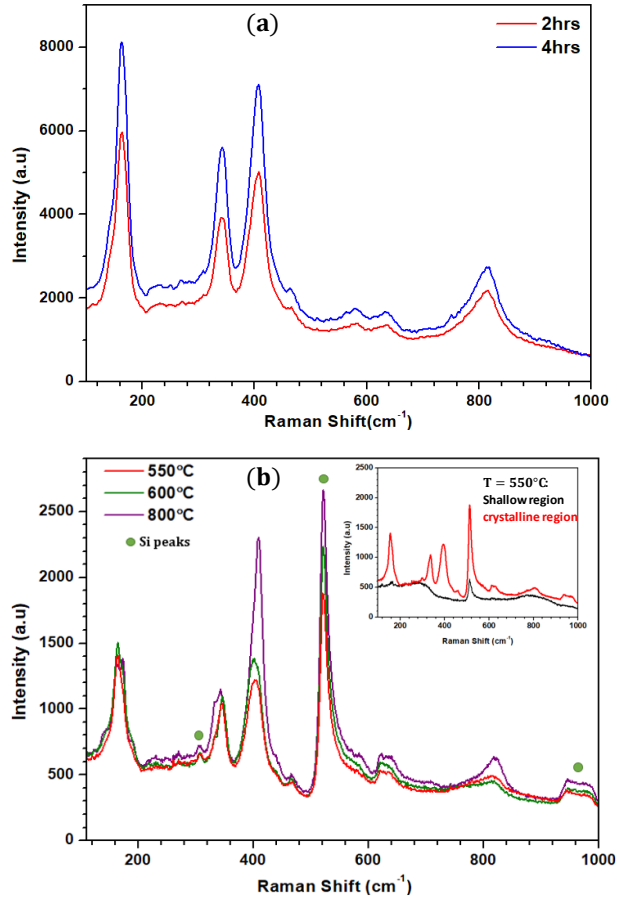}
\caption{Raman spectra showing the  effect of annealing (a) time (\NbO2 film on glass substrate) and (b) temperature on the film quality (\NbO2 film on Si substrate) . Inset: The Raman spectrum of the shallow non-crystalline region of the sample whose structure is shown in figure \ref{parameters2} (d).}
\label{comparison-temp}
\end{figure}

Raman Spectroscopy was also used to identify the crystalline phases in the sample. While the amorphous as-deposited films were featureless, annealed films showed peaks corresponding to the \NbO2 phase only. No other features corresponding to \Nb2O5 or other parasite phases were found on any of the films except for silicon peaks coming from the substrate. Raman spectra were used to quantify effects of annealing time and temperature. Figure \ref{comparison-temp} (a) shows the Raman spectra on a sample with a glass substrate after 2 and 4 hrs of annealing indicating that the \NbO2 signal to background ration is strengthened with increasing annealing time. The effect of annealing temperature is displayed in Figure \ref{comparison-temp} (b) where the strengthening of the \NbO2 signal to background ratio is even more remarkable, especially for the higher annealing temperature. These observations confirm the qualitative optical observations of Figure \ref{parameters2}.

%
%

\section{Transport Properties}

Sheet resistance measurements were made on both amorphous and crystalline \NbO2 thin films on glass substrates. The room temperature resistivity of \NbO2 in existing literature lacks consistency.  Values reported for crystalline  \NbO2  resistivity are for example,  200 $\Omega$.cm \cite{nakao}, 2.5 $\Omega$.cm \cite{wang}, $\sim$ 0.27 $\Omega$.cm \cite{resistivity-music}. It is well-known that amorphous \NbO2 is less resistive and a better thermal conductor than crystalline \NbO2. Nakao  et al. \cite{nakao} find that their amorphous films have a resistivity as low as 0.27 $\Omega$.cm, three orders of magnitude less than their crystalline films. Our crystalline films have a resistivity $\mathrm{\rho_{cry} \sim 54\, \Omega .cm}$ similar to that measured by Nakao et al. \cite{nakao}, but our amorphous films are only one order of magnitude less resistive ($\mathrm{\rho_{amo} \sim 4.5\, \Omega .cm}$).
The scatter in the results cited for resistivity in literature probably arises from two sources. Firstly, partial crystallization will influence the measured value if the films are partially crystallized. To illustrate this we measured the resistance $\mathrm{R_S}$ of three 33 nm thick samples deposited on glass which were first vacuum annealed at 600$\degree$C for 2 hrs (R=0.1 G$\Omega /\square $), 4 hrs (R=0.2  G$\Omega /\square$) and 6 hrs (R=0.3 G$\Omega /\square$). The increase in resistance with annealing time is in agreement with the earlier observations that the crystalline \NbO2 phase is more resistant than the amorphous one. Secondly, the stochiometry is also very important since any deviation from the \NbO2 stoichiometry will change resistivity, with a decrease in resistivity if the film is oxygen deficient (NbO is metallic)and an increase in resistivity if the film is too rich in oxygen (\Nb2O5 being a strong insulator). \\

\begin{figure}[t]
\hspace*{-0.8cm}
\centering
\includegraphics[scale=0.550]{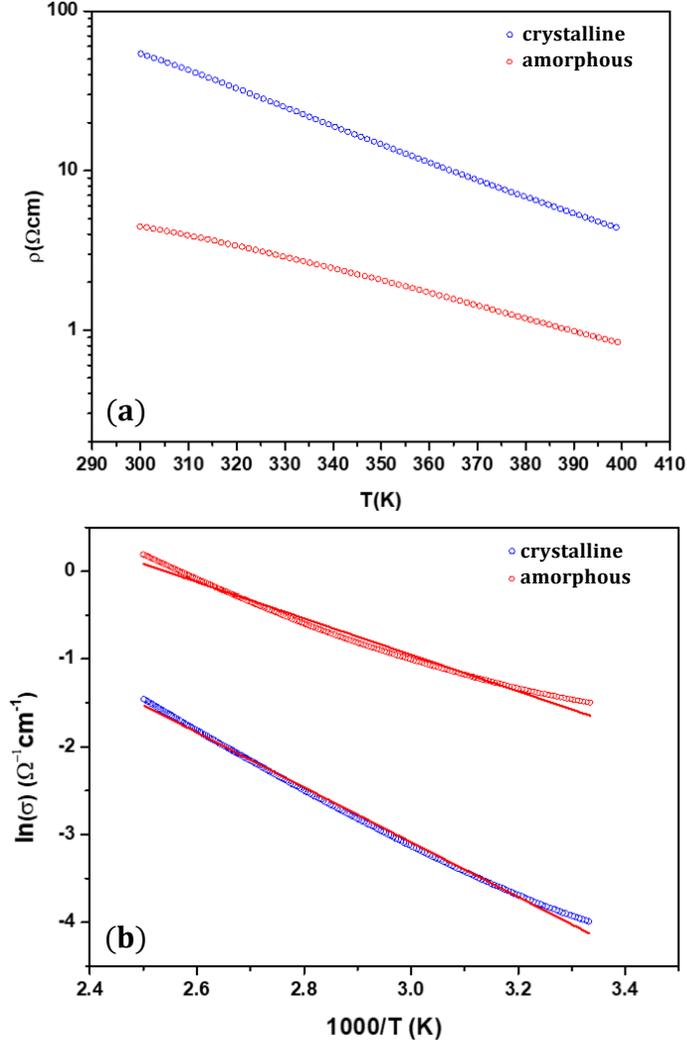}
\caption{(a) The resistivity of crystalline and amorphous \NbO2 films between 300 K and 400 K  deposited on glass substrates. (b) The corresponding   Arrhenius plots of the two samples from which the activation energy was extracted and found to be $\sim$ 0.27 eV and 0.18 eV for the crystalline and amorphous samples respectively. }
\label{amo-cry}
\end{figure} 

Finally, we measured the temperature dependent  resistivity  of an amorphous film  and a crystalline film  deposited on glass substrates between 300 K and 400 K as shown in Figure \ref{amo-cry} (a). The amorphous film is less resistive and as seen by the exponential decrease in resistivity with temperature, both display insulating behaviour each with a different activation energy. These activation energies are estimated from the linear fits in the Arrhenius plot of Figure \ref{amo-cry} (b) to be $\sim$ 0.27 eV and 0.18 eV for the crystalline and amorphous samples respectively.
%

\section{Conclusions}

We fabricated amorphous \NbO2 thin films (thickness between 33 and 330~nm) from a poly-crystalline target using RF-magnetron sputtering on three different substrates: glass, Si, SiO$_2$. The films were crystallized by annealing in vacuum at 600$\degree$C for films with glass substrates, and 600$\degree$C and above  for Si based films. We observed that a longer annealing time and higher temperature lead to better polycrystalline film quality. The room temperature resistivities of the crystalline and amorphous samples were extracted from the \Rs ~ measurements and were $\sim$ 54 $\Omega$.cm and $\sim$ 4.5 $\Omega$.cm respectively.

\begin{acknowledgments}
We acknowledge Benoit Baptiste and Ludovic Delbes for help with X-ray diffraction and Loic Becerra and Jean-Jacques Ganem for access to the clean room and vacuum furnace of the INSP. This work was supported by French state funds managed by the ANR within the Investissements d'Avenir programme under reference ANR-11-IDEX-0004-02, and more specifically within the framework of the Cluster of Excellence MATISSE led by Sorbonne Universit{\'e}s.

\end{acknowledgments}

\bibliographystyle{apsrev4-1}

\end{document}